\documentclass[lettersize,journal]{IEEEtran}
\usepackage{amsmath,amsfonts,amssymb,amsthm}
\usepackage{algorithmic}
\usepackage{algorithm}
\usepackage{array}
\usepackage[caption=false,font=footnotesize,labelfont=rm,textfont=rm]{subfig}
\usepackage{textcomp}
\usepackage{stfloats}
\usepackage{url}
\usepackage{capt-of}
\usepackage{verbatim}
\usepackage{graphicx}
\usepackage{cite}
\usepackage{xcolor}
\begin{document}

\title{Secure Wireless Communication via Polarforming}

\author{Jingze Ding, \IEEEmembership{Graduate Student Member, IEEE},
		Zijian Zhou, \IEEEmembership{Member, IEEE}, 
		Bingli Jiao, \IEEEmembership{Senior Member, IEEE},
		and Rui Zhang, \IEEEmembership{Fellow, IEEE}
\thanks{This work was supported in part by The Guangdong Provincial Key Laboratory of Big Data Computing, in part by the National Natural Science Foundation of China under Grant 62331022, in part by the Guangdong Major Project of Basic and Applied Basic Research under Grant 2023B0303000001, and in part by the High-Performance Computing Platform of Peking University. \textit{(Corresponding authors: Zijian Zhou; Rui Zhang.)}}
\thanks{Jingze Ding is with the School of Electronics, Peking University, Beijing 100871, China (e-mail: djz@stu.pku.edu.cn).}
\thanks{Zijian Zhou is with the School of Science and Engineering, The Chinese University of Hong Kong, Shenzhen 518172, China (e-mail: zijianzhou@link.cuhk.edu.cn).}
\thanks{Bingli Jiao is with the School of Computing and Artificial Intelligence, Fuyao University of Science and Technology, Fuzhou 350109, China. He is also with the School of Electronics, Peking University, Beijing 100871, China (e-mail: jiaobl@pku.edu.cn).}
\thanks{Rui Zhang is with the Department of Electrical and Computer Engineering, National University of Singapore, Singapore 117583 (e-mail: elezhang@nus.edu.sg).}
}
\maketitle

\begin{abstract}
Polarforming is a promising technique that enables dynamic adjustment of antenna polarization to mitigate depolarization effects commonly encountered during electromagnetic (EM) wave propagation. In this letter, we investigate the polarforming design for secure wireless communication systems, where the base station (BS) is equipped with polarization-reconfigurable antennas (PRAs) and can flexibly adjust the antenna polarization to transmit confidential information to a legitimate user in the presence of an eavesdropper. To maximize the achievable secrecy rate, we propose an efficient iterative algorithm to jointly optimize transmit beamforming and polarforming, where beamforming exploits spatial degrees of freedom (DoFs) to steer the transmit beam toward the user, while polarforming leverages polarization DoFs to align the polarization state of the EM wave received by the user with that of its antenna. Simulation results demonstrate that, compared to conventional fixed-polarization antenna (FPA) systems, polarforming can fully exploit the DoFs in antenna polarization optimization to significantly enhance the security performance of wireless communication systems.
\end{abstract}
\section{Introduction}
\IEEEPARstart{D}{ue} to the broadcast nature of wireless channels, security has emerged as a critical concern in future wireless communication systems. In conventional physical-layer security (PLS) systems, beamforming and interference management are essential methods to improve the signal reception at the legitimate user and degrade the signal quality at the eavesdropper \cite{survey,inter}. To enhance secure transmission, various techniques such as movable antenna (MA) \cite{ma1,ma2} and intelligent reflecting surface (IRS) \cite{irs1, irs2} have been explored by optimizing active and/or passive beamforming. Despite their effectiveness, these secure wireless communication systems have overlooked the polarization characteristics of electromagnetic (EM) waves. 

Unlike conventional beamforming, which exploits spatial degrees of freedom (DoFs) to prevent information leakage, the polarization domain offers additional DoFs for the design of secure wireless communication systems \cite{survey1}. Specifically, polarization diversity and multiplexing gains have been explored to strengthen security \cite{before1,before2,before3}. Nevertheless, these existing schemes rely on fixed-polarization antennas (FPAs), thus lacking the flexibility to fully exploit polarization DoFs. In practical propagation environments, EM waves may undergo depolarization due to factors such as reflection and scattering, resulting in the EM wave not retaining its initial polarization state when it reaches the receive antenna \cite{channel}.

To overcome this limitation, polarforming has emerged as an innovative solution that can dynamically reconfigure the antenna polarization in a cost-effective manner \cite{pf1,pf2,pf3,pf4}. Compared with most existing methods that reconfigure discrete polarization patterns through carefully designed antenna structures \cite{pra}, polarforming enables the antenna to be flexibly configured into arbitrary polarization states as required, by adjusting the phase differences among antenna elements within each antenna (e.g., a dual-polarized antenna with two co-located orthogonal elements, or a tri-polarized antenna with three such elements) through electronic phase shifting or mechanical antenna rotation, without modifying the physical antenna structure \cite{pf2}. In addition to conventional beamforming, polarforming can be integrated into secure wireless communication systems to improve robustness against eavesdropping, especially in scenarios with limited spatial DoFs. In particular, we can align the polarization state of the received signal with that of the legitimate user's antenna to maximize signal reception, while simultaneously inducing perfect polarization mismatching at the eavesdropper to suppress interception.

In light of the above, this letter investigates the polarforming design for secure wireless communication from a base station (BS) to a legitimate user in the presence of an eavesdropper. The BS is equipped with multiple polarization-reconfigurable antennas (PRAs), each containing two orthogonal antenna elements and capable of dynamically adjusting its polarization via a phase shifter. The user and the eavesdropper are each equipped with a single FPA. To maximize the achievable secrecy rate, we propose an efficient iterative algorithm to jointly optimize transmit beamforming and polarforming. Specifically, beamforming exploits spatial DoFs to steer the transmit beam toward the user, while polarforming leverages polarization DoFs to match/mismatch the polarization state of the received signal with that of the user's/eavesdropper's antenna. Simulation results demonstrate that the proposed system significantly improves the achievable secrecy rate compared to conventional FPA systems, owing to the polarforming gain provided by antenna polarization optimization.

The rest of this paper is organized as follows. Section \ref{sec2} introduces the system model and formulates the optimization problem. In Section \ref{sec3}, we propose an iterative algorithm to solve the optimization problem. Next, Section \ref{sec4} presents the simulation results and discussions. Finally, this paper is concluded in Section \ref{sec5}.

\textit{Notation:}  $a/A$, $\mathbf{a}$, and $\mathbf{A}$ denote a scalar, a vector, and a matrix, respectively. $\mathbf{A} \succeq \mathbf{0}$ indicates that $\mathbf{A}$ is a positive semidefinite matrix. ${\left(  \cdot  \right)^{T}}$, ${\left(  \cdot  \right)^{H}}$, $\left\|  \cdot  \right\|_2$, $\left\|  \cdot  \right\|_F$, $\left|  \cdot  \right|$, $\mathrm{tr}\left(   \cdot\right)   $, and $\mathrm{rank}\left(   \cdot\right)   $ denote the transpose, conjugate transpose, Euclidean norm, Frobenius matrix norm, absolute value, trace, and rank, respectively. $\odot$ denotes the Hadamard product. $\angle \mathbf{a} $ denotes the phase of complex vector $\mathbf{a}$. $\left[\mathbf{A} \right]_{ij} $ denotes the entry in the $i$-th row and $j$-th column of matrix $\mathbf{A}$. $\mathrm{blkdiag}\left(\mathbf{a}_1,\ldots, \mathbf{a}_N\right) $ denotes a block diagonal matrix composed of the blocks $\mathbf{a}_1,\ldots, \mathbf{a}_N$ along its diagonal. $\mathbb{C}^{M \times N}$ and $\mathbb{R}^{M \times N}$ are the sets for complex and real matrices of $M \times N$ dimensions, respectively. $\mathbf{I}_N$ is the identity matrix of order $N$. $\mathbf{0}_N$ denotes an $N$-dimensional vector with all elements equal to $0$. $\mathcal{CN}\left( 0, \sigma^2 \right) $ represents the circularly symmetric complex Gaussian (CSCG) distribution with mean zero and variance $\sigma^2$. $\sim$ and $\triangleq$ stand for ``distributed as'' and ``defined as'', respectively.
\section{System Model and Problem Formulation}\label{sec2}
\begin{figure}[!t]
	\centering
	\includegraphics[width=0.9\linewidth]{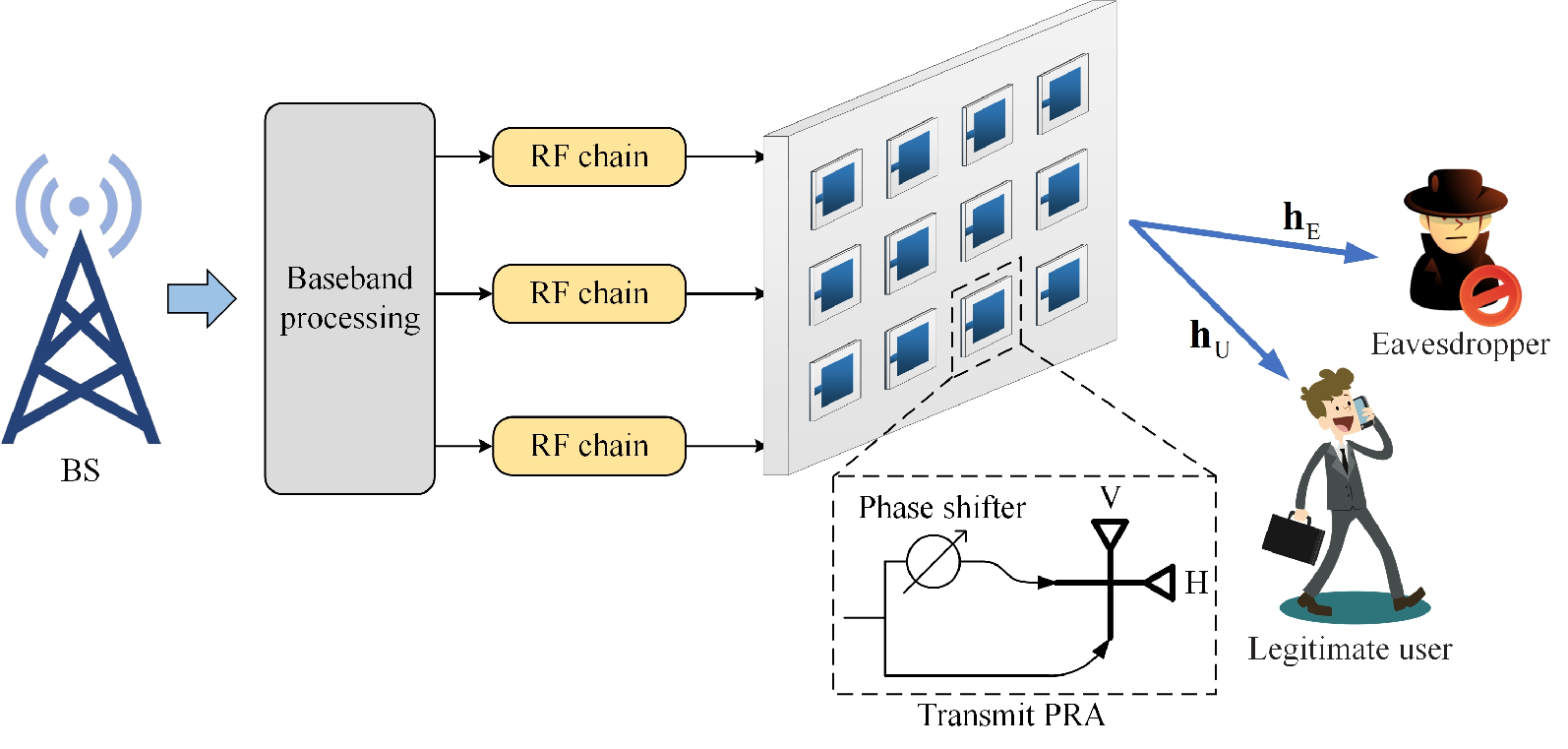}
	\caption{Illustration of the system model.}
	\label{sysmodel} 
\end{figure}
As shown in Fig. \ref{sysmodel}, we consider a secure wireless communication system where the BS intends to transmit confidential information to a legitimate user in the presence of an eavesdropper. The BS is equipped with a uniform planar array (UPA) with $N$ PRAs, each of which is connected to a dedicated radio frequency (RF) chain and consists of two orthogonal antenna elements, i.e., V-element for vertical polarization and H-element for horizontal polarization. Besides, a phase shifter is integrated into each PRA to control the phase difference between the signals on the V-element and the H-element (see Fig. \ref{sysmodel}), thereby dynamically controlling antenna polarization \cite{pf1}\footnote{Compared with the MA, which exploits the constructive or destructive superposition of multipath signals with different phases by moving the antenna physically, the phase-shifter-based PRA in this work reconfigures the antenna polarization by tuning the phase difference between two orthogonally polarized antenna elements within a single antenna at a fixed position \cite{pf4}.}. The user and the eavesdropper are each equipped with a single FPA.

Define $\theta_n$ as the phase shift of the $n$-th ($1 \le n \le N$) PRA. To characterize the antenna polarization, we define the transmit polarforming vector of the $n$-th PRA as a function of $\theta_n$ \cite{pf1}, i.e., 
\begin{equation} \label{f_n}
	\mathbf{f}\left(\theta_n \right) =\frac{1}{\sqrt{2}}\left[ 1,e^{\mathrm{j}\theta_n}\right]^T.
\end{equation}
Define ${{\mathbf{\Lambda}_{\mathrm{U},n}}} \in \mathbb{C}^{2 \times 2}$ and ${{\mathbf{\Lambda}_{\mathrm{E},n}}} \in \mathbb{C}^{2 \times 2}$ as the two-by-two polarized channel matrices from the $n$-th PRA to the user and the eavesdropper, respectively, which are given by \cite{channel}
\begin{equation}\label{polar_channel}
	{\mathbf{\Lambda}_{\mathrm{I},n}} = \frac{1}{{\sqrt {{\chi _\mathrm{I}} + 1} }}\left[ {\begin{array}{*{20}{c}}
			1&{\sqrt {{\chi _\mathrm{I}}} }\\
			{\sqrt {{\chi _\mathrm{I}}} }&1
	\end{array}} \right] \odot {{\tilde{\mathbf{H}}}_{\mathrm{I},n}},\quad \mathrm{I} \in \left\{\mathrm{U},\mathrm{E}\right\} ,
\end{equation}
where $\chi _\mathrm{I}$ denotes the inverse cross-polarization discrimination (XPD), which quantifies the degree of channel depolarization, and each element of ${{\tilde{\mathbf{H}}}_{\mathrm{I},n}}$ follows the CSCG distribution $\mathcal{CN}\left(0,{1}/{\sqrt{2}} \right) $ after normalization\footnote{The polarized channel model in \eqref{polar_channel} can be extended to account for practical propagation factors such as multipath effects and Doppler shifts \cite{pf4}.}. Let $\mathbf{g}_\mathrm{U} \in \mathbb{C}^{2 \times 1}$ and $\mathbf{g}_\mathrm{E} \in \mathbb{C}^{2 \times 1}$ denote the polarization vectors representing the antenna polarizations of the user and the eavesdropper, respectively. Define $\boldsymbol{\vartheta}\triangleq \left[\theta_1, \ldots, \theta _N \right]^T $ as the phase shift vector (PSV). Then, the channels from the BS to the user and the eavesdropper can be respectively expressed as ${\mathbf{h}_\mathrm{U}} \left( \boldsymbol{\vartheta}\right)   = {\left[ {\mathbf{g}_\mathrm{U}^H{\mathbf{\Lambda} _{\mathrm{U},1}}\mathbf{f}\left( {{\theta _1}} \right), \ldots ,\mathbf{g}_\mathrm{U}^H{\mathbf{\Lambda}_{\mathrm{U},N}}\mathbf{f}\left( {{\theta _N}} \right)} \right]^T}$ and ${\mathbf{h}_\mathrm{E}}\left( \boldsymbol{\vartheta}\right)  = {\left[ {\mathbf{g}_\mathrm{E}^H{\mathbf{\Lambda} _{\mathrm{E},1}}\mathbf{f}\left( {{\theta _1}} \right), \ldots ,\mathbf{g}_\mathrm{E}^H{\mathbf{\Lambda}_{\mathrm{E},N}}\mathbf{f}\left( {{\theta _N}} \right)} \right]^T}$, which are the functions of the PSV $\boldsymbol{\vartheta}$. To analyze the performance upper bound for the considered system, we assume that the channel state information (CSI) of all involved channels is perfectly known at the BS\footnote{It is worth noting that the CSI of the eavesdropping channel can be obtained when the eavesdropper is an active user in the system but is untrusted by the legitimate user \cite{irs1}, which constitutes an idealized assumption. The impact of imperfect CSI of the eavesdropper's channel on the considered system will be evaluated via simulations in Section \ref{sec4}.}, which can be acquired via the tensor-based channel estimation based on the pilot-based channel measurement\footnote{In the proposed polarforming system, channel estimation incurs additional delay and computational overhead due to multiple measurements under different polarization patterns, while the tuning accuracy of phase shifters affects the estimation performance. Hence, considering practical constraints in channel estimation and performance optimization for the polarforming system is an important topic worthy of further investigation.} \cite{est}.

To ensure secure wireless communication, the BS transmits confidential message $s$ with zero mean and unit power to the user via joint beamforming and polarforming, which exploit the spatial and polarization DoFs, respectively. Let $\mathbf{w} \in \mathbb{C}^{N \times 1}$ denote the beamforming vector. The received signals at the user and the eavesdropper can be respectively expressed as ${y_\mathrm{U}} = \mathbf{h}_\mathrm{U}^H\mathbf{w}s + {n_\mathrm{U}}$ and ${y_\mathrm{E}} = \mathbf{h}_\mathrm{E}^H\mathbf{w}s + {n_\mathrm{E}}$, where $n_\mathrm{U}$ and $n_\mathrm{E}$ denote the CSCG noises at the user and the eavesdropper with mean zero and variances $\sigma^2_\mathrm{U}$ and $\sigma^2_\mathrm{E}$, respectively. As such, the achievable secrecy rate in bits/second/Hz (bps/Hz) is given by ${R_{\sec }} = {[ {{{\log }_2}( {1 + {{{{| {\mathbf{h}_\mathrm{U}^H\mathbf{w}} |}^2}}}/{{{\sigma_\mathrm{U} ^2}}}} ) - {{\log }_2}( {1 + {{{{| {\mathbf{h}_\mathrm{E}^H\mathbf{w}} |}^2}}}/{{{\sigma_\mathrm{E} ^2}}}} )} ]^ + }$, where $\left[ x\right]^+\triangleq\max\left( x,0\right) $.

In this letter, we aim to maximize the secrecy rate $R_{\sec }$ by jointly optimizing the transmit beamforming vector $\mathbf{w}$ and the BS's PSV $\boldsymbol{\vartheta}$. Accordingly, the secrecy rate maximization problem is formulated as
\begin{subequations}
	\label{max1}
	\begin{align}
		& \mathop {\mathrm{max} }\limits_{\mathbf{w},\boldsymbol{\vartheta}} \quad R_{\sec } \\
		&\hspace{0.3em}\mathrm{s.t.} \quad {\left\| \mathbf{w} \right\|_2^2} \le {P_{\max }}, \label{c1} \\
		&\hspace{0.3em}\hspace{2.3em}  \theta_n \in [0, 2\pi] ,\quad \forall 1 \le n \le N, \label{c2}
	\end{align}
\end{subequations}
where $P_{\max }$ is the power budget of the BS. In the following analysis, we omit the operator $\left[ x\right]^+$ because the optimal value of problem \eqref{max1} must be non-negative \cite{irs1}. Problem \eqref{max1} is difficult to solve because its objective function is non-concave with respect to either $\mathbf{w}$ or $\boldsymbol{\vartheta}$. Thus, in the next section, we propose an efficient iterative algorithm to solve problem \eqref{max1}.
\section{Proposed Solution}\label{sec3}
In problem \eqref{max1}, it is worth noting that constraint \eqref{c1} is only related to $\mathbf{w}$ and constraint \eqref{c2} is only related to $\boldsymbol{\vartheta}$. Therefore, problem \eqref{max1} can be efficiently solved by alternately optimizing $\mathbf{w}$ and $\boldsymbol{\vartheta}$ in an iterative manner. In the following, we develop dedicated algorithms for optimizing $\mathbf{w}$ with given $\boldsymbol{\vartheta}$ (beamforming optimization) and optimizing $\boldsymbol{\vartheta}$ with given $\mathbf{w}$ (polarforming optimization).
\subsection{Beamforming Optimization}
Define $\mathbf{A}_\mathrm{U} \triangleq \frac{1}{\sigma _\mathrm{U}^2}{\mathbf{h}_\mathrm{U}}\mathbf{h}_\mathrm{U}^H \in \mathbb{C}^{N \times N}$ and $\mathbf{A}_\mathrm{E} \triangleq\frac{1}{\sigma _\mathrm{U}^2} {\mathbf{h}_\mathrm{E}}\mathbf{h}_\mathrm{E}^H \in \mathbb{C}^{N \times N}$. With given $\boldsymbol{\vartheta}$, problem \eqref{max1} is simplified to
\begin{equation}\label{max2}
	\mathop {\mathrm{max} }\limits_{\mathbf{w}} \quad \frac{{ {\mathbf{w}^H}\mathbf{A}_\mathrm{U}\mathbf{w}+1}}{{ {\mathbf{w}^H}\mathbf{A}_\mathrm{E}\mathbf{w}+1}} \quad \mathrm{s.t.} \quad \eqref{c1}.
\end{equation}
The optimal solution to problem \eqref{max2} is \cite{irs1}
\begin{equation}\label{beam}
	\mathbf{w} = \sqrt{P_\mathrm{max}}\mathbf{u}_\mathrm{max},
\end{equation}
where $\mathbf{u}_\mathrm{max} \in \mathbb{C}^{N \times 1}$ is the normalized eigenvector associated with the largest eigenvalue of the matrix ${\left( {\mathbf{A}_\mathrm{E} + \frac{1}{P_\mathrm{max}}{\mathbf{I}_N}} \right)^{ - 1}}\left( {\mathbf{A}_\mathrm{U} + \frac{1}{P_\mathrm{max}}{\mathbf{I}_N}} \right)$.
\subsection{Polarforming Optimization}
Define $\mathbf{b}_{\mathrm{I},n}\triangleq\mathbf{\Lambda}^H_{\mathrm{I},n}\mathbf{g}_\mathrm{I} \in \mathbb{C}^{2 \times 1}$ and $\mathbf{B}_\mathrm{I}\triangleq\mathrm{blkdiag}\left(\mathbf{b}_{\mathrm{I},1}, \ldots\mathbf{b}_{\mathrm{I},N} \right) \in \mathbb{C}^{2N \times N}$, $\mathrm{I} \in \left\{\mathrm{U},\mathrm{E}\right\} $. Since the phase shift $\theta_n$ in \eqref{f_n} is only used to adjust the phase difference between the signals at the two antenna elements, we can rewrite the channels from the BS to the user and the eavesdropper as
\begin{equation} \label{channel_I}
	{\mathbf{h}_\mathrm{I}}\left( \boldsymbol{\varphi}  \right) = \frac{1}{{\sqrt 2 }}\mathbf{B}_\mathrm{I}^H\tilde {\mathbf{f}}\left( \boldsymbol{\varphi} \right), \quad \mathrm{I} \in \left\{\mathrm{U},\mathrm{E}\right\},
\end{equation}
where $\boldsymbol{\varphi}=\left[\varphi_1,\ldots,\varphi_{2N} \right]^T$, $\tilde {\mathbf{f}}\left( \boldsymbol{\varphi}  \right) = {\left[ {{e^{\mathrm{j}{\varphi _1}}}, \ldots ,{e^{\mathrm{j}{\varphi _{2N}}}}} \right]^T}$, and $\varphi_{2n}-\varphi_{2n-1}=\theta_n$, $\forall 1 \le n \le N$. With \eqref{channel_I} and the given $\mathbf{w}$, problem \eqref{max1} can be reformulated as
\begin{subequations}
	\label{max3}
\begin{align}
	& \mathop {\mathrm{max} }\limits_{\boldsymbol{\varphi}} \quad \frac{{{{\tilde{\mathbf{f}}}^H\left( \boldsymbol{\varphi}\right) }{\mathbf{D}_\mathrm{U}}\tilde{\mathbf{f}}\left( \boldsymbol{\varphi}\right) + 1}}{{{{\tilde{\mathbf{f}}}^H\left( \boldsymbol{\varphi}\right)}{\mathbf{D}_\mathrm{E}}\tilde{\mathbf{f}\left( \boldsymbol{\varphi}\right)} + 1}} \\
	&\hspace{0.3em}\mathrm{s.t.} \quad \varphi_l \in \left[ 0,2\pi\right] ,\quad \forall 1 \le l \le 2N, \label{c3}
\end{align}
\end{subequations}
where ${\mathbf{D}_\mathrm{I}} = \frac{1}{{2\sigma _\mathrm{I}^2}}{\mathbf{B}_\mathrm{I}}\mathbf{w}{\mathbf{w}^H}\mathbf{B}_\mathrm{I}^H \in \mathbb{C}^{2N \times 2N}$, $\mathrm{I} \in \left\{\mathrm{U},\mathrm{E}\right\} $. To reformulate problem \eqref{max3} into a more tractable form, we introduce $2N$ auxiliary matrices, denoted as $\mathbf{E}_l \in \mathbb{R}^{2N \times 2N}$ ($1 \le l \le 2N$). The ($i,j$)-th element of $\mathbf{E}_l$ satisfies 
\begin{equation}
	{\left[ {{\mathbf{E}_l}} \right]_{ij}} = \left\{ {\begin{array}{*{20}{c}}
			{1,\quad i = j = l,}\\
			{\hspace{0.2em}0,\quad \mathrm{otherwise}.}
	\end{array}} \right.
\end{equation}
Therefore, problem \eqref{max3} can be reformulated as
\begin{subequations}
	\label{max4}
\begin{align}
	& \mathop {\mathrm{max} }\limits_{\tilde{\mathbf{f}}} \quad \frac{{{{\tilde{\mathbf{f}}}^H }{\mathbf{D}_\mathrm{U}}\tilde{\mathbf{f}} + 1}}{{{{\tilde{\mathbf{f}}}^H}{\mathbf{D}_\mathrm{E}}\tilde{\mathbf{f}} + 1}} \\
	&\hspace{0.3em}\mathrm{s.t.} \quad  \tilde{\mathbf{f}}^H\mathbf{E}_l\tilde{\mathbf{f}}=1,\quad \forall 1 \le l \le 2N.\label{c4}
\end{align}
\end{subequations}However, since the objective function in problem \eqref{max4} is non-concave with respect to $\tilde{\mathbf{f}}$ and constraint \eqref{c4} consists of $2N$ non-convex quadratic equality constraints, finding the optimal solution to problem \eqref{max4} remains challenging. Thus, an efficient algorithm is proposed to obtain an approximate solution to problem \eqref{max4} as follows.

Define $\mathbf{F}\triangleq\tilde{\mathbf{f}}\tilde{\mathbf{f}}^H \in \mathbb{C}^{2N \times 2N}$. Then, it follows that $\mathrm{rank}\left( \mathbf{F}\right) =1$. By relaxing the rank-one constraint $\mathrm{rank}\left( \mathbf{F}\right) =1$, problem \eqref{max4} can be rewritten as
\begin{subequations}
	\label{max5}
\begin{align}
	& \mathop {\mathrm{max} }\limits_{\mathbf{F}\succeq \mathbf{0}} \quad \frac{{\mathrm{tr}\left( {{\mathbf{D}_\mathrm{U}}\mathbf{F}} \right) + 1}}{{\mathrm{tr}\left( {{\mathbf{D}_\mathrm{E}}\mathbf{F}} \right) + 1}} \\
	&\hspace{0.3em}\mathrm{s.t.} \quad  \mathrm{tr}\left( {{\mathbf{E}_l}\mathbf{F}} \right)=1,\quad \forall 1 \le l \le 2N.\label{c5}
\end{align}
\end{subequations}
To transform the objective function in problem \eqref{max5} into a non-fractional form, we apply the Charnes-Cooper transformation \cite{irs1} and set $\mu  = \frac{1}{{\mathrm{tr}\left( {{\mathbf{D}_\mathrm{E}}\mathbf{F}} \right) + 1}}$ and $\mathbf{S}  = \mu \mathbf{F}$. Then, problem \eqref{max5} is transformed into
\begin{subequations}
	\label{max6}
\begin{align}
	& \mathop {\mathrm{max} }\limits_{\mathbf{S}\succeq \mathbf{0},  \mu \ge 0} \quad {\mathrm{tr}\left( {{\mathbf{D}_\mathrm{U}}\mathbf{S}} \right) + \mu} \\
	&\hspace{1em}\mathrm{s.t.} \quad  \mathrm{tr}\left( {{\mathbf{D}_\mathrm{E}}\mathbf{S}} \right) + \mu=1, \label{c6} \\
	&\hspace{1em}\hspace{2.3em} \mathrm{tr}\left( {{\mathbf{E}_l}\mathbf{S}} \right)=\mu ,\quad \forall 1 \le l \le 2N,\label{c7}  
\end{align}
\end{subequations}
which is a convex semidefinite programming (SDP) problem and can be optimally solved using CVX \cite{cvx}. To address the relaxed constraint $\mathrm{rank}\left( \mathbf{F}\right) =1$, we use the Gaussian randomization method detailed in \cite{irs3} to obtain an approximate solution to problem \eqref{max4}. Finally, by setting $\boldsymbol{\varphi}=\angle{\tilde{\mathbf{f}}\left(\boldsymbol{\varphi} \right) }$ and $\theta_n=\varphi_{2n}-\varphi_{2n-1}$, $\forall 1 \le n \le N$, we can recover the optimized $\boldsymbol{\vartheta}$.
\begin{algorithm}[!t]
	\caption{Proposed algorithm for solving problem \eqref{max1}}
	\label{alg1}
	\footnotesize
	\renewcommand{\algorithmicrequire}{\textbf{Initialization:}}
	\renewcommand{\algorithmicensure}{\textbf{Output:}}
	\begin{algorithmic}[1]
		\REQUIRE Set iteration index $k = 0$, $\boldsymbol{\vartheta}^{\left( 0\right) }=\mathbf{0}_N$, and error tolerance $0 \le \epsilon \ll 1$.
		\ENSURE $\mathbf{w}$, $\boldsymbol{\vartheta}$.
		\REPEAT
		\STATE Set $ k=k+1$;
		\STATE With given $\boldsymbol{\vartheta}^{\left( k-1\right) }$, update $\mathbf{w}^{\left( k\right) }$ by \eqref{beam};
		\STATE With given $\mathbf{w}^{\left( k\right)}$, solve problem \eqref{max6} and recover the optimized $\boldsymbol{\vartheta}^{\left( k\right) }$;
		\UNTIL {Increase of $R_\mathrm{sec}$ is less than $\epsilon$}
		\RETURN $\mathbf{w}=\mathbf{w}^{\left( k\right)}$ and $\boldsymbol{\vartheta}=\boldsymbol{\vartheta}^{\left( k\right) }$.
	\end{algorithmic}
\end{algorithm}
\subsection{Overall Algorithm}
The overall algorithm for solving problem \eqref{max1} is detailed in Algorithm \ref{alg1}, and its convergence is guaranteed because the objective value $R_\mathrm{sec}$ in problem \eqref{max1} is non-decreasing over iterations and is upper-bounded due to limited communication resources. In addition, the computational complexity of Algorithm \ref{alg1} mainly comes from lines 3 and 4. In line 3, the computational complexity for solving problem \eqref{max2} based on \eqref{beam} is of order $\mathcal{O}( N^3) $ \cite{irs1}. In line 4, the worst-case complexity for solving the SDP problem \eqref{max6} is of order $\mathcal{O}( ( 2N) ^{6.5}) $ \cite{cvx1}. Therefore, the computational complexity of Algorithm \ref{alg1} is approximately of order $\mathcal{O}( K( N^{3}+( 2N) ^{6.5}) ) $, where $K$ denotes the number of iterations.
\section{Simulation Results} \label{sec4}
In this section, we present simulation results to demonstrate the effectiveness of polarforming in enhancing secure wireless communication performance. We assume that the average noise powers at the user and the eavesdropper are identical, i.e., $\sigma^2_\mathrm{U}=\sigma^2_\mathrm{E}=\sigma^2$, and define the signal-to-noise ratio (SNR) as $\gamma=P_\mathrm{max}/\sigma^2$. We adopt the polarized channel model in \eqref{polar_channel}, where the inverse XPD of both the user's and eavesdropper's channels is set to $\chi_\mathrm{U}=\chi_\mathrm{E}=0.2$. Unless otherwise specified, the user and the eavesdropper are each equipped with a single linearly polarized antenna, i.e., $\mathbf{g}_\mathrm{U}=\left[ 1,0\right]^T$ and $\mathbf{g}_\mathrm{E}=\left[ 1,0\right]^T$, respectively \cite{pf1}. The number of BS's antennas is set to $N=4$, the SNR is set to $\gamma=0$ dB, and the error tolerance is set to $\epsilon =10^{-5}$. Besides, the following benchmark schemes are defined for performance comparison: 1) \textit{Upper bound}: An upper bound of the secrecy rate in Algorithm \ref{alg1} is obtained from the objective value of the relaxed problem \eqref{max6}. 2) \textit{MRT}: The transmit beamforming vector is designed using the maximum ratio transmission (MRT) method based on the legitimate user's channel only, i.e., $\mathbf{w}=\sqrt{P_\mathrm{max}}\mathbf{h}_\mathrm{U}/\left\|\mathbf{h}_\mathrm{U}\right\|_2$. 3) \textit{FPA}: The antenna polarization of the BS is fixed to circular polarization, i.e., $\mathbf{f}\left(\theta_n \right) =\frac{1}{\sqrt{2}}\left[ 1,\mathrm{j}\right]^T$, $\forall 1 \le n \le N$.

\begin{figure*}[!t]
	\centering
	\begin{minipage}[t]{0.31\textwidth}
		\centering
		\includegraphics[width=\linewidth]{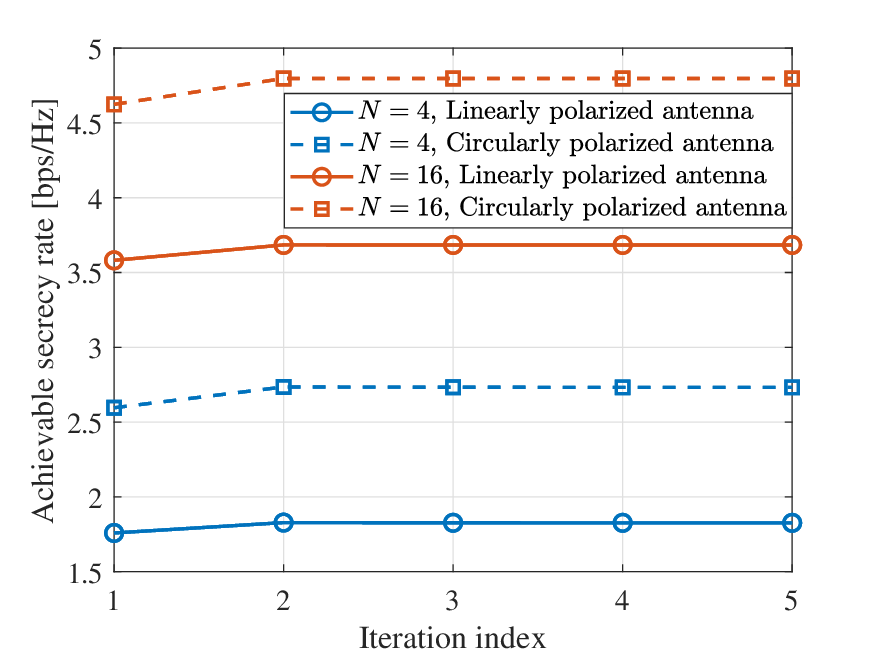}
		\captionof{figure}{Convergence behavior of Algorithm \ref{alg1}.}
		\label{conv}
	\end{minipage}
	\begin{minipage}[t]{0.31\textwidth}
		\centering
		\includegraphics[width=\linewidth]{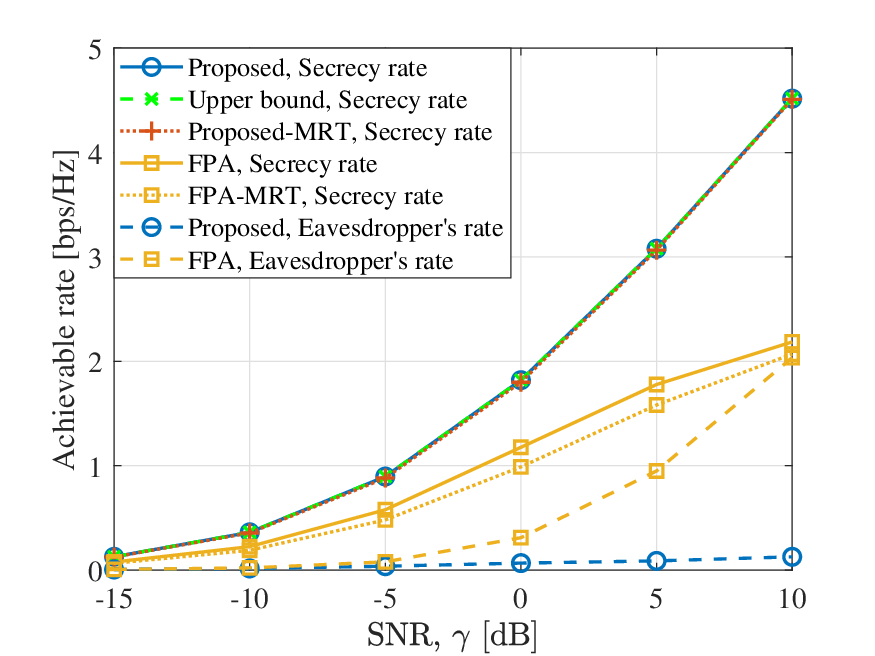}
		\captionof{figure}{Secrecy rate and eavesdropper's rate versus SNR.}
		\label{SNR}
	\end{minipage}
	\begin{minipage}[t]{0.31\textwidth}
		\centering
		\includegraphics[width=\linewidth]{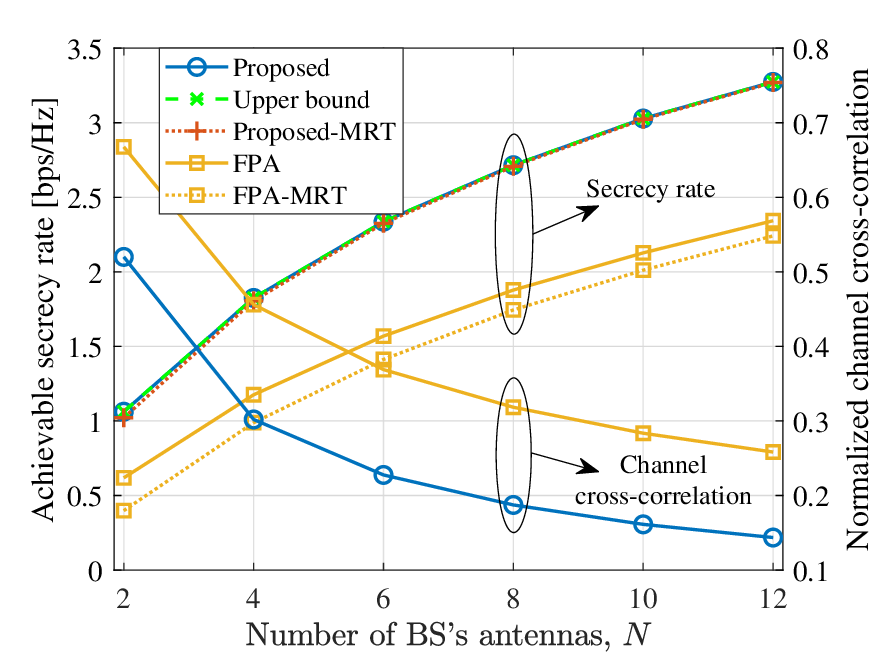}
		\captionof{figure}{Secrecy rate and channel cross-correlation versus number of BS's antennas.}
		\label{N}
	\end{minipage}
\end{figure*}
First, Fig. \ref{conv} shows the convergence behavior of Algorithm \ref{alg1}, where both the user and the eavesdropper are equipped with a linearly polarized antenna (solid lines) or a circularly polarized antenna (dashed lines). The circularly polarized antenna is implemented by orthogonally co-locating two linearly polarized antenna elements with a phase difference of $\pi/2$, and thus $\mathbf{g}_\mathrm{U}=\left[ 1,\mathrm{j}\right]^T$ and $\mathbf{g}_\mathrm{E}=\left[ 1,\mathrm{j}\right]^T$. In all cases, Algorithm \ref{alg1} achieves convergence within three iterations, thereby validating its rapid convergence performance. In addition, we can see that Algorithm \ref{alg1} already achieves favorable performance after the first iteration, which makes the subsequent iterations yield only marginal performance improvements. This is because the beamforming optimization problem is optimally solved in closed form, and the polarforming optimization problem is efficiently solved based on the SDP formulation. Moreover, since the circularly polarized antenna comprises more antenna elements than the linearly polarized antenna, polarforming can exploit more polarization DoFs to achieve better security performance when the user and the eavesdropper are equipped with the circularly polarized antenna.

Next, Fig. \ref{SNR} illustrates the secrecy rates and the eavesdropper's rates of different schemes versus the SNR $\gamma$. The eavesdropper's rate is defined as ${{\log }_2}( {1 + {{| {\mathbf{h}_\mathrm{E}^H\mathbf{w}} |}^2/\sigma ^2}} )$. It can be observed that the secrecy rate and the eavesdropper's rate both increase with $\gamma$, as the higher transmit power enhances the signal reception quality at the user and the eavesdropper. Compared to the FPA scheme, the eavesdropper's rate of the proposed scheme is significantly lower. This is because the BS adaptively adjusts its antenna polarization so that the polarization state of the received signal at the eavesdropper is mismatched with its antenna polarization, thereby effectively preventing information leakage. Also note that with transmit polarforming, the MRT scheme and the proposed scheme achieve nearly the same secrecy rate, which is mainly due to the low polarized fading channel correlation between the user and the eavesdropper, rendering the MRT and the proposed beamforming design in \eqref{beam} both very effective. Thus, in this scenario, MRT can be a viable low-complexity alternative for beamforming optimization in the proposed system.

Moreover, Fig. \ref{N} presents the secrecy rate and the channel cross-correlation versus the number of BS's antennas $N$, where the normalized channel cross-correlation is defined as $\frac{{\left| {\mathbf{h}_\mathrm{U}^H{\mathbf{h}_\mathrm{E}}} \right|}}{{{{\left\| {{\mathbf{h}_\mathrm{U}}} \right\|}_2}{{\left\| {{\mathbf{h}_\mathrm{E}}} \right\|}_2}}}$. We can see that the proposed scheme consistently outperforms the FPA scheme, and the performance gap between them increases with $N$. This is because polarforming reduces the polarized channel correlation between the user and the eavesdropper (see Fig. \ref{N}) by introducing polarization matching and mismatching for them, respectively, which makes transmit beamforming more effective for secrecy communication, even for the suboptimal MRT scheme. In addition, the available spatial and polarization DoFs are expanded as $N$ increases. Furthermore, it is observed that the secrecy rate of the proposed scheme is very close to its upper bound, which indicates that the approximate solution obtained by the Gaussian randomization method is nearly optimal.

\begin{figure}[!t]
	\centering
	\begin{minipage}[t]{0.24\textwidth}
		\centering
		\includegraphics[width=\linewidth]{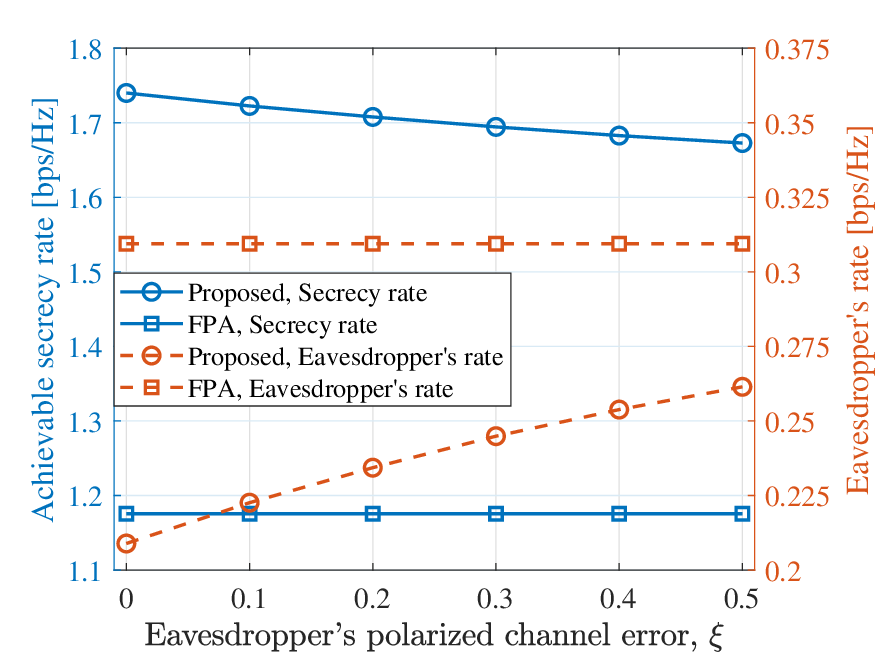}
		\captionof{figure}{Secrecy rate and eavesdropper's rate versus normalized variance of eavesdropper's polarized channel error.}
		\label{err}
	\end{minipage}
	\begin{minipage}[t]{0.24\textwidth}
		\centering
		\includegraphics[width=\linewidth]{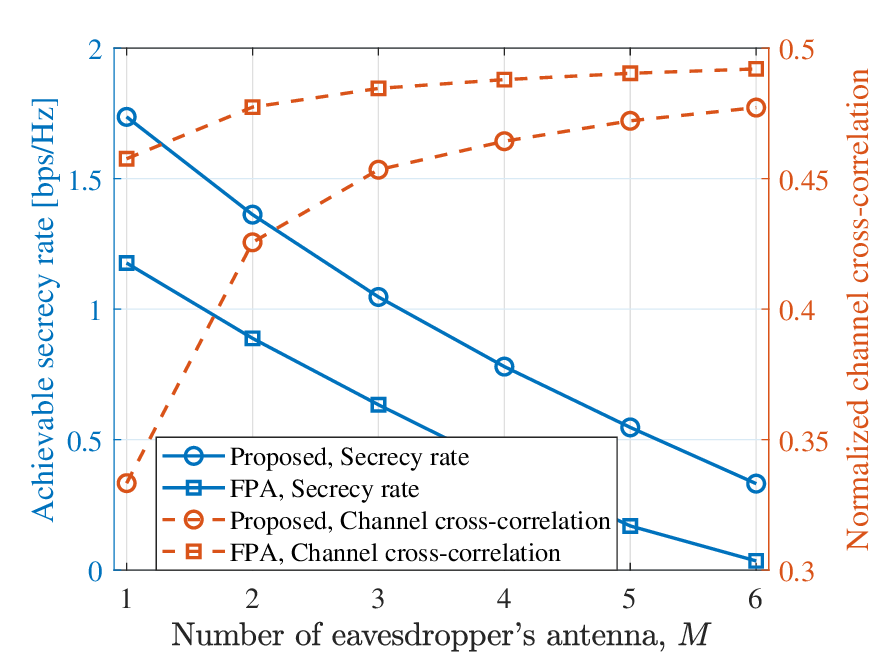}
		\captionof{figure}{Secrecy rate and channel cross-correlation versus number of eavesdropper's antennas.}
		\label{M}
	\end{minipage}
\end{figure}

Furthermore, Fig. \ref{err} evaluates the impact of the eavesdropper's polarized channel error on the performance of the proposed scheme. The eavesdropper's polarized channel error is defined as the difference between its actual and estimated polarized channel matrices, i.e., $\frac{{{{\left[ {{\mathbf{\Lambda} _{\mathrm{E},n}}} \right]}_{ij}} - {{[ {{{\hat {\mathbf{\Lambda}} }_{\mathrm{E},n}}} ]}_{ij}}}}{{\left| {{{\left[ {{\mathbf{\Lambda}_{\mathrm{E},n}}} \right]}_{ij}}} \right|}} \sim \mathcal{CN}\left( {0,\xi } \right)$, $\forall 1 \le i,j \le 2$, $\forall 1 \le n \le N$, where ${\hat {\mathbf{\Lambda}} }_{\mathrm{E},n}$ is the estimated polarized channel and $\xi$ represents the normalized variance of the polarized channel error. To focus on investigating the effect of imperfect eavesdropper's polarized channel estimation on the antenna polarization optimization, the transmit polarforming is first optimized by solving problem \eqref{max6} based on the estimated channel, and then the transmit beamforming is updated by \eqref{beam} based on the actual channel. Therefore, the performance of the FPA scheme does not vary with $\xi$. As shown in Fig. \ref{err}, when $\xi$ increases from 0 to 0.5, the secrecy rate of the proposed scheme slightly decreases by 3.86\%, and its eavesdropper's rate slightly increases by 2.01\%. As a result, the proposed scheme still yields a superior performance to the conventional FPA scheme for large values of $\xi$, thereby indicating its high robustness against the estimation error of the eavesdropper's polarized channel.

Finally, Fig. \ref{M} investigates the impact of the number of eavesdropper's antennas on the secrecy performance of the proposed system. Specifically, an eavesdropper equipped with multiple receive antennas can maximize its eavesdropping SNR by employing maximum ratio combining (MRC). Without loss of generality, we assume that the eavesdropper is equipped with a UPA consisting of $M$ linearly polarized antennas, and thus the channel between the BS and the eavesdropper is extended to $\mathbf{H}_\mathrm{E} \in \mathbb{C}^{N \times M}$, as detailed in \cite[Section II-A]{pf1}. Accordingly, the achievable secrecy rate and the normalized channel cross-correlation are redefined as ${[ {{{\log }_2}( {1 + {{| {\mathbf{h}_\mathrm{U}^H\mathbf{w}} |}^2/\sigma ^2}} ) - {{\log }_2}( {1 + {|| {\mathbf{H}_\mathrm{E}^H\mathbf{w}} ||_2^2/\sigma ^2}} )} ]^ + }$ and $\frac{{{{\left\| {\mathbf{h}_\mathrm{U}^H{\mathbf{H}_\mathrm{E}}} \right\|}_2}}}{{{{\left\| {{\mathbf{h}_\mathrm{U}}} \right\|}_2}{{\left\| {{\mathbf{H}_\mathrm{E}}} \right\|}_F}}}$, respectively. To investigate the impact of multiple antennas equipped by the eavesdropper on the performance of polarforming designed for a single-antenna eavesdropper, we first optimize the transmit polarforming based on the eavesdropper's channel with a single antenna by solving problem \eqref{max6}, and then update the transmit beamforming based on the eavesdropper's channel with $M$ antennas by \eqref{beam}, where $\mathbf{A}_\mathrm{E}$ in problem \eqref{max2} is replaced by $\tilde{\mathbf{A}}_\mathrm{E} =\frac{1}{\sigma^2} {\mathbf{H}_\mathrm{E}}\mathbf{H}_\mathrm{E}^H \in \mathbb{C}^{N\times N}$. We can see in Fig. \ref{M} that the secrecy rate decreases as $M$ increases because more eavesdropper's antennas enhances the channel correlation between the user and the eavesdropper, thereby increasing the risk of information leakage. However, the proposed scheme can still achieve significant performance gains over the conventional FPA scheme in the presence of a multi-antenna eavesdropper, which demonstrates the effectiveness of polarforming against the eavesdropper with higher spatial DoFs.
\section{Conclusion}\label{sec5}
This letter proposed a new secure wireless communication system, where the BS can dynamically adjust antenna polarization via polarforming to suppress the leakage of confidential information to the eavesdropper. We investigated the joint optimization of transmit beamforming and polarforming to fully exploit the spatial and polarization DoFs, respectively. It was shown by simulations that the proposed system can effectively utilize antenna polarization optimization to enhance security performance compared to conventional FPA systems. Future work can be extended to investigate the robust design of polarforming for secure communication systems in the presence of multiple eavesdroppers and imperfect CSI of eavesdropping channels.


\begin{thebibliography}{1}
\bibliographystyle{IEEEtran}

\bibitem{survey}
X. Chen, D. W. K. Ng, W. H. Gerstacker, and H.-H. Chen, ``A survey on multiple-antenna techniques for physical layer security,'' \textit{IEEE Commun. Surveys Tuts.}, vol. 19, no. 2, pp. 1027--1053, 2nd Quart. 2017.

\bibitem{inter}
Z. Wei, C. Masouros, F. Liu, S. Chatzinotas, and B. Ottersten, ``Energy- and cost-efficient physical layer security in the era of IoT: The role of interference,'' \textit{IEEE Commun. Mag.}, vol. 58, no. 4, pp. 81--87, Apr. 2020.

\bibitem{ma1}
G. Hu, Q. Wu, K. Xu, J. Si, and N. Al-Dhahir, ``Secure wireless communication via movable-antenna array,'' \textit{IEEE Signal Process. Lett.}, vol. 31, pp. 516--520, Jan. 2024.

\bibitem{ma2}
J. Ding, Z. Zhou, and B. Jiao, ``Movable antenna-aided secure full-duplex multi-user communications,'' \textit{IEEE Trans. Wireless Commun.}, vol. 24, no. 3, pp. 2389--2403, Mar. 2025.

\bibitem{irs1}
M. Cui, G. Zhang, and R. Zhang, ``Secure wireless communication via intelligent reflecting surface,'' \textit{IEEE Wireless Commun. Lett.}, vol. 8, no. 5, pp. 1410--1414, Oct. 2019.

\bibitem{irs2}
X. Yu, D. Xu, Y. Sun, D. W. K. Ng, and R. Schober, ``Robust and secure wireless communications via intelligent reflecting surfaces,'' \textit{IEEE J. Sel. Areas Commun.}, vol. 38, no. 11, pp. 2637--2652, Nov. 2020.

\bibitem{survey1}
C. Guo, F. Liu, S. Chen, C. Feng, and Z. Zeng, ``Advances on exploiting polarization in wireless communications: Channels, technologies, and applications,'' \textit{IEEE Commun. Surveys Tuts.}, vol. 19, no. 1, pp. 125--166, 1st Quart. 2017.

\bibitem{before1}
F. Wang and X. Zhang, ``Secure resource allocations for polarization-enabled multiple-access cooperative cognitive radio networks with energy harvesting capability,'' \textit{IEEE Trans. Wireless Commun.}, vol. 22, no. 8, pp. 4990--5004, Aug. 2023.

\bibitem{before2}
Y. Ding and V. Fusco, ``Polarization distortion as a means for securing wireless communication,'' in \textit{Proc. 8th Eur. Conf. Antennas Propag. (EuCAP)}, The Hague, Netherlands, 2014, pp. 1203--1207.

\bibitem{before3}
C. Qian and Z. Jiang, ``A wireless secure communication scheme based on polarization scrambling,'' in \textit{Proc. 25th Wireless Opt. Commun. Conf. (WOCC)}, Chengdu, China, 2016, pp. 1--4.

\bibitem{channel}
Y. He, X. Cheng, and G. L. Stuber, ``On polarization channel modeling,'' \textit{IEEE Wireless Commun.}, vol. 23, no. 1, pp. 80--86, Feb. 2016.

\bibitem{pf1}
Z. Zhou, J. Ding, C. Wang, B. Jiao, and R. Zhang, ``Polarforming for wireless communications: Modeling and performance analysis,'' 2024, \textit{arXiv:2409.07771}.

\bibitem{pf2}
J. Ding, Z. Zhou, X. Shao, B. Jiao, and R. Zhang, ``Polarforming for wireless networks: Opportunities and challenges,'' 2025, \textit{arXiv:2505.20760}.

\bibitem{pf3}
X. Shao, R. Zhang, H. Zhou, Q. Jiang, C. Zhou, W. Zhuang, and X. Shen, ``Polarforming antenna enhanced sensing and communication: Modeling and optimization,'' \textit{IEEE J. Sel. Areas Commun.}, early access, Sep. 16, 2025, doi: 10.1109/JSAC.2025.3610487.

\bibitem{pf4}
Z. Zhou, J. Ding, and R. Zhang, ``Polarforming design for movable antenna systems,'' \textit{IEEE Wireless Commun. Lett.}, early access, Oct. 17, 2025, doi: 10.1109/LWC.2025.3622756.

\bibitem{pra}
S. -L. Chen, Y. Liu, H. Zhu, D. Chen, and Y. J. Guo, ``Millimeter-wave cavity-backed multi-linear polarization reconfigurable antenna,'' \textit{IEEE Trans. Antennas Propag.}, vol. 70, no. 4, pp. 2531--2542, Apr. 2022. 

\bibitem{est}
J. He, Y. Wang, T. Shu, and T.-K. Truong, ``Polarization, angle, and delay
estimation for tri-polarized systems in multipath environments,'' \textit{IEEE Trans. Wireless Commun.}, vol. 21, no. 8, pp. 5828--5841, Aug. 2022.

\bibitem{cvx}
S. Boyd and L. Vandenberghe, \textit{Convex Optimization}. Cambridge, U.K.: Cambridge Univ. Press, 2004.

\bibitem{irs3}
Q. Wu and R. Zhang, ``Intelligent reflecting surface enhanced wireless network via joint active and passive beamforming,'' \textit{IEEE Trans. Wireless Commun.}, vol. 18, no. 11, pp. 5394--5409, Nov. 2019.

\bibitem{cvx1}
Z. -Q. Luo and W. Yu, ``An introduction to convex optimization for communications and signal processing,'' \textit{IEEE J. Sel. Areas Commun.}, vol. 24, no. 8, pp. 1426--1438, Aug. 2006.
\end{thebibliography}
\end{document}